\documentclass[showkeys,showpacs,preprintnumbers,amsmath,amssymb]{revtex4}

\usepackage{graphicx}
\usepackage{dcolumn}
\usepackage{bm}

\begin{document}

\title{Boundary Shape and Casimir Energy}

\author{H. Ahmedov}
\email{hagi@gursey.gov.tr} \affiliation{ Feza G\"ursey Institute, P.
K. 6  \c Cengelk\" oy, 34684 Istanbul, Turkey. Tel:+90-216-308 94 30
}

\author{I.H. Duru}
\email{ihduru@galois.iyte.edu.tr}
 \affiliation{ Izmir  Institute of
Technology, 35430, Izmir, Turkey }

\date{\today}

\begin{abstract}
Casimir energy changes are investigated for geometries   obtained by
small but arbitrary deformations of a given  geometry  for which the
vacuum energy is already known  for the massless scalar field. As a
specific case, deformation of a spherical shell  is studied. From
the deformation of the sphere  we show that the Casimir energy is a
decreasing function of the surface to volume ratio. The decreasing
rate is higher for  less smooth deformations.
\end{abstract}
\pacs{03.65.-w, 03.70.+k} \keywords{Casimir energy, boundary
variations and  approximations. }


\maketitle

\section{Introduction}
Casimir energies are known for several cavities in several spatial
dimensions for  electromagnetic  or massless scalar fields. Exact
Casimir energy calculations are available for  rectangular prisms
\cite{Par}, for spherical shell  \cite{Sp}, for  cylindrical region
\cite{Cyl}, for  a pyramidal cavity   and for  a conical
 cavity \cite{Haji}. All these geometries  have definite boundary wall
shapes.  For example the prisms are all with right angular wedges,
the sphere is the perfect one; and, the pyramid and the cone are of
very special types. The obvious reason of these restrictions is the
fact that these are the regions for which one can calculate the
exact field modes with the required boundary conditions. Limited
number of examples ( all with rigid walls ) do of course not give
much hint about the dependence of the Casimir energy on the shapes
of the regions.
\\
\\
There are to our knowledge two approaches in dealing with rather
arbitrary geometries. First one is the proximity force approximation
\cite{5}. It is applicable to two body systems which are close to
each other. It employs the parallel plate modes in every cylindrical
region of infinitesimal base between the bodies and then integrates
over these regions \cite{6}. Second approach is the one called
multiple scattering expansion. It formulates the vacuum energy for
the electromagnetic field in terms of the successive scattering from
the conducting boundaries \cite{7}. In this approach between two
successive scatterings free Green function is employed. The method
enables one to investigate the connections between the divergencies
and the geometrical details of the boundaries. Employment of the
free Green functions however from one scattering to the next one is
not of much practical value in compact space regions: Specially for
more curved boundaries one may need to consider large number of
scatterings to approximate the exact Green function. On the other
hand it may be more reasonable to employ the exact Green function (
if it is already known ) of the compact region ( instead of the free
one ) if the region under consideration is sufficiently close to the
original region. Variations around the exactly solvable geometries
may give some hint about the dependence of the vacuum energies on
the shape of the boundaries.
\\
\\
In present work we try to investigate the shape dependence of the
Casimir energy for massless scalar field by calculating the effect
of the small but arbitrary deformations of a given geometry  for
which  we already know the vacuum energies. We use Green function to
formulate the perturbation theory  around the exact solution of the
region with boundary $S$. Suppose $G_\omega^S$ is the exact Green
function for the massless scalar field confined to the region in
$S$, and $\beta$ is the small deformation of $S$. Converting a
boundary problem into an integral equation  we arrive at the
perturbation series
$$
G_\omega^{\tilde{S}}=G_\omega^S+\beta G_\omega^{1S}+\beta^2
G_\omega^{2S}+\cdots
$$
where $G_\omega^{jS}$ is the correction to the original Green
function $G_\omega^S$ resulting from the $j$ times reflections from
the deformed boundary $\tilde{S}$. One reflection gives information
about the size of the new boundary. To get information about the
shape dependence we need to take into account at least two
reflections.
\\
\\
Having in hand the Green function  we can  construct the zeta
function which is useful  tool in  Casimir energy calculation. The
zeta function of the system can be expressed in terms of the heat
kernel coefficients which are functionals of the geometrical
invariants of the boundaries \cite{9}. For arbitrary geometries
these coefficients are too complicated and therefore are not
available for practical purposes \cite{8}. From this point of view
we think that the perturbation around a known geometry may be an
effective approach.
\\
\\
For massless fields which are the only fields for which the Casimir
energy is meaningful, there is no unique way of getting rid of the
infinities if the heat kernel coefficient $a_2$ of the zeta function
expansion is not zero \cite{Most}. The situation can be improved if
one considers the whole space for when one sums the zeta functions
of the in and out regions   $a_2$ coefficients cancel each other. Of
course for such cancelation the boundaries should be free of sharp
corners.
\\
\\
In the coming section we briefly review the zeta function approach
to the vacuum energy calculations and the regularization scheme
which we employ in the our work.
\\
\\
In Section III we present  the general formulation of the Casimir
energy contribution  of  small deformations of the boundaries.
\\
\\
In Section IV deformation of the sphere is discussed.
\\
\\
In Section V we  analyze the dependence of the energy on the shape
of the boundary.
\\
\\
Details of the involved calculations are given in the Appendices.

\section{ A Brief Review of the Zeta Function Method}

Formally the calculation of the Casimir energy is reduced to a
treatment of a sum over all one particle energy eigenvalues
\begin{equation}
    E=\frac{1}{2}\sum_{\lambda\in \Lambda} \sqrt{E_\lambda}
\end{equation}
This sum is divergent and regularization is needed. For the scalar
field confined in a compact three dimensional region with the
Dirichlet boundary condition the zeta function
\begin{equation}
   \zeta(z) =\sum_{n=1}^\infty E_n^{-z}
\end{equation}
is well defined by Weyl theorem for $Re z > 3/2$ \cite{8}. By the
analytic continuation it is possible to define this function on the
whole complex plane. This may be done by using the representation
\begin{equation}
   \zeta(z)=\frac{1}{\Gamma (z)}\int_0^\infty dt t^{z-1} K(t)
\end{equation}
with the heat-kernel
\begin{equation}
   K(t)=\sum_{n=1}^\infty e^{-tE_n}.
\end{equation}
For $t\rightarrow \infty $ the integral is well behaved. Possible
poles arise  from $t\rightarrow 0$ behavior of the heat kernel
\cite{9}
\begin{equation}
    K(t)\sim \sum_{n=0,1/2,1,\dots}^\infty a_n t^{n-3/2}.
\end{equation}
Splitting the integral as $\int_0^1dt +\int_1^\infty dt$ we arrive
at
\begin{equation}
   Res ( \zeta (z)\Gamma (z))\mid_{z=3/2-n}=a_n
\end{equation}
where $a_n$ are heat kernel coefficients which depent  on the
geometry of boundary which confines the scalar field. When the
coefficient $a_2$ is nonzero the value of the zeta function at
$z=-1/2$ which defines the Casimir energy becomes infinite. For
massless scalar field there is no renormalization condition to get
rid of this pole in a unique way \cite{Most}. One needs additional
considerations to apply the zeta function method to the calculation
of the vacuum energy. The extrinsic curvature of the sphere will
have opposite sign when viewed from inside or outside. $a_2$ heat
-kernel coefficient depends on an odd power of extrinsic curvature.
Two of them add to cancel each other when we approach the surface
both from in and out region of the ball.  This does not hold only
for the spherical shell but is a general property for boundaries of
an arbitrary shape. This cancelation of poles occurs only for
infinitely thin boundaries. Once a finite thickness is introduced
the absolute value of the extrinsic curvature at the inner and outer
side of the boundary is different and divergencies do not cancel
each other.
\\
\\
In the present work we try to investigate the shape dependence of
the Casimir energy for the massless scalar field by calculating the
effect of the small but arbitrary smooth  deformations of the
boundary of given regions for which we already know the vacuum
energies. We restrict our attention to the deformations of the
spherical shell. Using the scattering theory in the spherical
coordinates one arrives at the zeta functions inside
\begin{equation}\label{inside}
  \zeta_{in} (z)=\frac{\sin\pi z}{\pi}\sum_{l=0}^\infty
  (2l+1)\int_0^\infty d\omega \omega^{-2z} \frac{d}{d\omega}\ln
  (\omega^{-l-1/2}I_{l+1/2}(\omega))
\end{equation}
and outside the ball \cite{8, Most}
\begin{equation}\label{outside}
  \zeta_{out} (z)=\frac{\sin\pi z}{\pi}\sum_{l=0}^\infty
  (2l+1)\int_0^\infty d\omega \omega^{-2z} \frac{d}{d\omega}\ln
  (\omega^{l+1/2}K_{l+1/2}(\omega))
\end{equation}
The zeta function in the whole space
\begin{equation}
      \zeta (z)= \frac{1}{2}(\zeta_{in}(z)+\zeta_{out}(z))
\end{equation}
or
\begin{equation}\label{whole}
       \zeta (z)=\frac{\sin\pi z}{\pi}\sum_{l=0}^\infty
  (l+1/2)\int_0^\infty d\omega \omega^{-2z} \frac{d}{d\omega}\ln
  (I_{l+1/2}(\omega)K_{l+1/2}(\omega))
\end{equation}
 is well defined at $z=-1/2$. To find this value we use the uniform
 asymptotic expansions for Bessel functions \cite{Stegun}
 \begin{equation}
   K_\nu (\nu x)= \sqrt{\frac{\pi}{2\nu}}
   \frac{e^{-\nu\eta}}{(1+x^2)^{1/4}}(1+\sum_{k=1}^\infty (-)^k
   \frac{u_k(t)}{\nu^k})
 \end{equation}
and
 \begin{equation}
   I_\nu (\nu x)= \frac{1}{\sqrt{2\pi\nu}}
   \frac{e^{\nu\eta}}{(1+x^2)^{1/4}}(1+\sum_{k=1}^\infty
   \frac{u_k(t)}{\nu^k})
 \end{equation}
where
\begin{equation}
 t=\frac{1}{\sqrt{1+x^2}}, \ \ \eta=\sqrt{1+x^2}+\ln \frac{x}{1+\sqrt{1+x^2}}
\end{equation}
and coefficients $u_k(t)$ satisfy the recurrence relation
\begin{equation}
    u_{k+1}(t)=\frac{1}{2 }t^2(1-t^2)u^\prime_k(t)+\frac{1}{8}\int_0^t
    d\tau (1-5\tau^2)u_k(\tau)
\end{equation}
with the initial condition $u_0(t)=1$. The uniform asymptotic
expansions imply
\begin{equation}
\ln I_\nu(\nu x)=\sum_{-1}^\infty \frac{X_k(t)}{\nu^k}
\end{equation}
and
\begin{equation}\label{UNE}
\ln K_\nu(\nu x)=\sum_{-1}^\infty (-)^k\frac{X_k(t)}{\nu^k}
\end{equation}
where the first four terms of $X_n(t)$s are
\begin{eqnarray}
   X_{-1}&=&\frac{1}{t}+\ln \frac{t}{1+t}, \nonumber \\
   X_0 &=& \frac{1}{2}\ln t, \nonumber \\
   X_1 &=& \frac{t}{8}-\frac{5t^3}{24}, \nonumber \\
   X_2 &=&\frac{t^2}{16}-\frac{3t^4}{8}+\frac{5t^6}{16}.
\end{eqnarray}
The zeta function in the whole space becomes
\begin{equation}
       \zeta (-1/2)=\frac{2}{\pi} \sum_{m=0}^\infty \zeta_0 (2m-2)  \int_0^\infty dx  X_{2m}(t)
\end{equation}
where
\begin{equation}
  \zeta_0(s)=\sum_{n=0}^\infty \frac{1}{(1/2+n)^s}
\end{equation}
is the Riemann zeta function which vanishes for  $z=0$ and $z=-2$
\cite{Grad}. Therefore the expansion (18) converges. The vacuum
energy for the massless scalar field in the spherical shell of
radius $R$ is \cite{cognola}
\begin{equation}\label{Casimirsphere}
    E_{sph}=\frac{1}{2R}\zeta(-1/2)\simeq\frac{\alpha}{R}, \ \ \  \alpha\simeq 0,003.
\end{equation}

\section{Contribution of  Small Boundary Deformations to the Vacuum Energy}

In this section we use the Green function representation of the zeta
function for the massless scalar field in the three dimensional
space vanishing on a surface $S$
\begin{equation}\label{zetagreen}
  \zeta^S (z)= \frac{\sin(\pi z)}{\pi}\int_0^\infty  d\omega  \omega^{-2z+1} \int_{R^3} d^3 \vec{x}  G_\omega^S(\vec{x},\vec{x})
\end{equation}
where
\begin{equation}
   G_\omega^S(\vec{x},\vec{y})=\{\begin{array}{c}
                                 G_\omega^{in S}(\vec{x},\vec{y}), \ \  \vec{x},\vec{y} \in \Omega_{in}\\
                                G_\omega^{out S}(\vec{x},\vec{y}), \ \  \vec{x},\vec{y} \in \Omega_{out}
                               \end{array}
\end{equation}
is the Green function in $R^3$ satisfying the boundary problem
\begin{equation}\label{Green}
    (-\Delta + \omega^2 )  G^S_\omega(\vec{x},\vec{y})=\delta (\vec{x} -\vec{y}), \
    \  \        G^S_\omega(\vec{x},\vec{y} )=0, \ \   \vec{x}\in S.
\end{equation}
Here $G_\omega^{in S}$ ( $G_\omega^{outS}$ ) is the Green function
in the in-region $\Omega_{in}$ ( the out-region $\Omega_{out}$ ). To
have a well defined  integral over $R^3$ in (\ref{zetagreen}) one
considers  a ball of radius $L$ and then let it go to infinite.
 Divergent terms in powers of L
will correspond to the infinite vacuum oscillations of  the free
Minkowski space.  The variation
\begin{equation}\label{Var}
    \delta\zeta(z)=\zeta^{\tilde{S}}(z)-\zeta^{S}(z)
\end{equation}
resulting from the deformation of the boundary $S$ is then given by
\begin{equation}\label{DSpectral}
\delta\zeta(z) =  \frac{\sin(\pi z)}{\pi}\int_0^\infty  d\omega
\omega^{-2z+1} \int_{R^3} d^3 \vec{x} \delta
G_\omega(\vec{x},\vec{x} )
\end{equation}
where
\begin{equation}
   \delta G_\omega(\vec{x},\vec{y})=
   G^{\tilde{S}}_\omega(\vec{x},\vec{y})-G^S_\omega(\vec{x},\vec{y}).
\end{equation}
Due to   (\ref{Green})  it satisfies  the wave equation
\begin{equation}
(-\Delta + \omega^2 )  \delta G_\omega(\vec{x},\vec{y})=0
\end{equation}
and the boundary condition
\begin{equation}
   \delta G_\omega(\vec{x},\vec{y} )= -G^{S}_\omega(\vec{x}, \vec{y}), \ \ \ \  \vec{x}\in \tilde{S}.
\end{equation}
The above boundary problem is equivalent to the integral equation
\cite{Vlad}
\begin{equation}\label{integ}
  G^{\tilde{S}}_\omega(\vec{x},\vec{y})= G^{S}_\omega(\vec{x},\vec{y})-\int_{\tilde{S}} d\tilde{s}
   \frac{\partial G^{\tilde{S}}_\omega(\vec{x},\vec{v})}{\partial m(\vec{v})}
G^S_\omega(\vec{v},\vec{y}),
\end{equation}
where
\begin{equation}
    \frac{\partial}{\partial m(\vec{v})}=\vec{ m} (\vec{v})
    \frac{\partial }{\partial \vec{v}}
\end{equation}
is the derivation along the unit vector $\vec{m} (\vec{v})$ normal
to the wall   $\tilde{S}$ at a point $\vec{v}$. In the parametric
representation $\vec{v}=\vec{v}(\tau)$, $\tau=(\tau^1, \tau^2)$ the
integration measure on $\tilde{S}$ is
\begin{equation}
   d\tilde{s}= \sqrt{\mid \tilde{g}\mid} d^2\tau
\end{equation}
where  $\mid \tilde{g}\mid$ is  the determinant of the induced
metric
\begin{equation}
    \tilde{g}_{ab}=(\frac{\partial \vec{v}}{\partial \tau^a},\frac{\partial \vec{v}}{\partial \tau^b})
\end{equation}
with $(\cdot,\cdot)$ being  the scalar product in the three
dimensional space.
\\
\\
The solution of the integral equation (\ref{integ}) up to the second
order ( which we can also interpret as the second reflection )
 is
\begin{equation}
 G^{\tilde{S}}_\omega(\vec{x},\vec{y})= G^{S}_\omega(\vec{x},\vec{y})-\int_{\tilde{S}} d\tilde{s}
   \frac{\partial G^{S}_\omega(\vec{x},\vec{v})}{\partial m(\vec{v})}
G^S_\omega(\vec{v},\vec{y})+\int_{\tilde{S}}
d\tilde{s}\int_{\tilde{S}} d\tilde{s}
   \frac{\partial G^{S}_\omega(\vec{x},\vec{v})}{\partial m(\vec{v})}\frac{\partial G^{S}_\omega(\vec{v},\vec{v^\prime})}{\partial m(\vec{v\prime})}
G^S_\omega(\vec{v^\prime},\vec{y}).
\end{equation}
The property
\begin{equation}
    \int_{R^3}
    d^3\vec{x}G_\omega^S(\vec{z},\vec{x})G_\omega^S(\vec{x},\vec{z^\prime})=-\frac{\partial}{\partial
    \omega^2}G_\omega^S(\vec{z},\vec{z^\prime})
\end{equation}
allows us to integrate explicitly  the perturbation solution over
the three dimensional spatial space to get
\begin{equation}
 G^{\tilde{S}}_\omega= G^{S}_\omega + \frac{1}{2}\frac{\partial}{\partial \omega^2}\int_{\tilde{S}} d\tilde{s}
   \frac{\partial G^{S}_\omega (\vec{v},\vec{v})}{\partial m(\vec{v})}
-\frac{\partial}{\partial \omega^2}\int_{\tilde{S}}
d\tilde{s}\int_{\tilde{S}} d\tilde{s}
   \frac{\partial G^{S}_\omega(\vec{v^\prime},\vec{v})}{\partial m(\vec{v})}\frac{\partial G^{S}_\omega(\vec{v},\vec{v^\prime})}{\partial
   m(\vec{v\prime})}.
\end{equation}
We  consider deformations of the boundary $S$ along the unit vector
$\vec{n}(z)$ normal to the surface $S$ at a point $\vec{z}$:
\begin{equation}\label{deformation}
   \vec{v}= \vec{z}- \beta \vec{n}(\vec{z})f (\vec{z})
\end{equation}
where $\beta$ is the dimensionless deformation parameter of the
surface $S$. This deformation formula  implies
\begin{equation}
    \tilde{g}_{ab} =  g_{ab}-\beta (\frac{\partial \vec{z}}{\partial \tau^a},\frac{\partial f \vec{n}}{\partial \tau^b})+(a\rightarrow
    b))
\end{equation}
where $g_{ab}$ is the metric tensor on $S$. Using $\delta \sqrt{\mid
g\mid}=1/2\sqrt{\mid g\mid} g^{ab}\delta g_{ab}$ we arrive at the
variation of the integration measure
\begin{equation}
   d\tilde{s}=ds-\beta g^{ab}(\frac{\partial \vec{z}}{\partial \tau^a},\frac{\partial (f \vec{n})}{\partial
   \tau^b})ds.
\end{equation}
which together with  the Taylor expansion
\begin{equation}\label{taylor}
   G^S_\omega(\vec{x},\vec{v})=-\beta f(\vec{z})\frac{\partial G^S_\omega(\vec{x},\vec{z})}
    {\partial n(\vec{z})}+\frac{\beta^2}{2}f^2(\vec{z})\frac{\partial^2 G^S_\omega(\vec{x},\vec{z})}{\partial
    n^2(\vec{z})}+\cdots
\end{equation}
implies
\begin{eqnarray}\label{Varfinal1}
 \delta G_\omega &=&  - \frac{\beta}{2}\frac{\partial}{\partial \omega^2}\int_{S}
 ds   f(\vec{z})\frac{\partial^2 G^{S}_\omega(\vec{z},\vec{z})}{\partial
 n^2(\vec{z})}+\frac{\beta^2}{2}\frac{\partial}{\partial \omega^2}\int_{S}
 ds   g^{ab}(\frac{\partial \vec{z}}{\partial \tau^a},\frac{\partial (f
\vec{n})}{\partial
   \tau^b})f^2(\vec{z})\frac{\partial^2 G^{S}_\omega(\vec{z},\vec{z})}{\partial
 n^2(\vec{z})} \nonumber \\
&+& \frac{\beta^2}{4}\frac{\partial}{\partial \omega^2}\int_{S}
 ds   f^2(\vec{z})\frac{\partial^3 G^{S}_\omega(\vec{z},\vec{z})}{\partial
 n^3(\vec{z})}
-\beta^2\frac{\partial}{\partial \omega^2}\int_{S} ds\int_{S}
ds^\prime
   f(\vec{z})f(\vec{z^\prime})(\frac{\partial^2 G^{S}_\omega(\vec{z^\prime},\vec{z})}{\partial n(\vec{z})\partial
   n(\vec{z^\prime})})^2
\end{eqnarray}
 Up to the second order in $\beta$  the zeta function  variation
(\ref{Var}) becomes
\begin{equation}\label{Varfinal2}
\delta \zeta(z) = \frac{\sin (\pi z)}{\pi}\int_0^\infty d\omega
\omega^{-2z+1} \delta G_\omega.
\end{equation}

\section{ Deformation of the Spherical Shell}
 The in and out-Green function  for the massless scalar field vanishing on the  sphere of the radius $R$
are
\begin{equation}
    G_\omega^{in} (r,\vec{n}; r^\prime
    \vec{n^\prime})=
    -\frac{1}{4\pi\sqrt{rr^\prime}}\sum_{l=0}^\infty( 2l+1)P_l((\vec{n},
    \vec{n^\prime})) \frac{ I_{l+1/2}(\omega r)( K_{l+1/2}(\omega R)I_{l+1/2}(\omega r^\prime)-I_{l+1/2}(\omega R)K_{l+1/2}(\omega r^\prime)}
                         {I_{l+1/2}(\omega R)}
\end{equation}
and
\begin{equation}
    G_\omega^{out} (r,\vec{n}; r^\prime
    \vec{n^\prime})=
    -\frac{1}{4\pi\sqrt{rr^\prime}}\sum_{l=0}^\infty( 2l+1)P_l((\vec{n},
    \vec{n^\prime})) \frac{( K_{l+1/2}(\omega R)I_{l+1/2}(\omega r)-I_{l+1/2}(\omega R)K_{l+1/2}(\omega r)) K_{l+1/2}(\omega r^\prime)}
                         {K_{l+1/2}(\omega R)}
\end{equation}
where $0\leq r\leq r^\prime \leq R$,  $P_l (x)$ is the Legendre
polynomial and   $\vec{n}$ is  the unit vector normal to the sphere
( see Appendix B ). The  derivative normal to the sphere is
$\frac{\partial}{\partial n(\vec{z})}=\frac{\partial}{\partial r}$.
We  have the derivatives of the following type
\begin{equation}
  \frac{\partial^2 G_\omega^{S^2}(r,\vec{n}; r^\prime
    \vec{n^\prime})}{\partial r \partial r^\prime }\mid_{r,r^\prime
    =R}=-\frac{\omega}{2\pi R}\sum_{l=0}^\infty( 2l+1)P_l((\vec{n},
    \vec{n^\prime})) T_{l+1/2}(\omega )
\end{equation}
where
\begin{equation}\label{SPF}
    T_{l+\frac{1}{2}}(\omega)=\frac{1}{2}
   \frac{d}{d\omega}
   \ln (I_{l+\frac{1}{2}}(\omega)K_{l+\frac{1}{2}}(\omega))
\end{equation}
is the spectral function in the whole space. Inserting the above
type terms in (\ref{Varfinal1}),  (\ref{Varfinal2}) becomes
\begin{eqnarray}\label{zetain}
\delta \zeta (z) &=& -\frac{2z\sin(\pi z) }{\pi R}\sum_{l=0}^\infty
(2l+1)\int_0^\infty d\omega \omega^{-2z} T_{l+\frac{1}{2}}(\omega)
 \int d\Omega (\frac{\beta}{4\pi R}f(\vec{n}) + \frac{\beta^2}{4\pi R^2}f^2(\vec{n}) )  \nonumber \\
         &+&  \frac{8z\sin(\pi z)}{\pi R}\sum_{l,l^\prime=0}^\infty
         (l+\frac{1}{2})(l^\prime+\frac{1}{2})D_{ll^\prime}
         \int_0^\infty   d\omega  \omega^{1-2z}
         (T_{l+\frac{1}{2}}(\omega) -
         T_{l^\prime+\frac{1}{2}}(\omega))^2
\end{eqnarray}
where
\begin{equation}
   D_{ll^\prime}=\frac{\beta^2}{16\pi^2R^2}\int d\Omega\int d\Omega^\prime P_l((\vec{n}, \vec{n^\prime})) P_{l^\prime}
   ((\vec{n}, \vec{n^\prime}))f(\vec{n}) f(\vec{n^\prime}),
\end{equation}
$d\Omega =d\phi d\theta \sin\theta$ is the integration measure on
the sphere.  The first term in (\ref{zetain}) at $z=-1/2$ is
proportional to the vacuum energy of the massless scalar field
confined in the spherical region of the radius R. Therefore the
variation of the vacuum energy is then
\begin{equation}\label{energyvar}
\delta E = E_{sph} \int d\Omega (\frac{\beta}{4\pi R}f(\vec{n}) +
\frac{\beta^2}{4\pi R^2}f^2(\vec{n}) )+
        \frac{2}{\pi R}\sum_{l,l^\prime=0}^\infty
         (l+\frac{1}{2})(l^\prime+\frac{1}{2})D_{ll^\prime}
         \int_0^\infty   d\omega  \omega^2
         (T_{l+\frac{1}{2}}(\omega) -
         T_{l^\prime+\frac{1}{2}}(\omega))^2
\end{equation}
The expansion (\ref{expansion}) and the addition formula (\ref{add})
imply
\begin{equation}\label{DDD}
   D_{ll^\prime}=\beta^2\sum_{J=0}^\infty \sum_{M=-J}^J  K^J_{ll^\prime}
   \frac{\mid f^J_M\mid ^2}{(2J+1)^2}
\end{equation}
where $K^J_{ll^\prime}$ are the Clebsch Gordon coefficients
(\ref{KG}) and $f^J_M$ are the expansion coefficients of the
deformation function $f$ in the spherical harmonics:
\begin{equation}\label{sphericalharmonics}
 f(\vec{n})=\sum_{J=0}^\infty \sum_{M=-J}^J f^J_M Y^J_M(\vec{n}).
\end{equation}
Using (\ref{DDD}) we may represent (\ref{energyvar}) as
\begin{equation}
   \delta E= E_{sph}[ \frac{\beta}{4\pi R}  \int d\Omega f(\vec{n})
    +\frac{\beta^2}{4\pi R^2}\int d\Omega ( f^2(\vec{n})+f(\vec{n})\hat{ H} f(\vec{n}))]
\end{equation}
where  $\hat{H}$ is an  energy operator
\begin{equation}\label{Hoper}
    \hat{H}Y^J_M(\vec{n})=H(J) Y^J_M(\vec{n})
\end{equation}
with e-values
\begin{equation}\label{CasIII}
    H(J) = \frac{1}{\pi \alpha } \sum_{l=J}^\infty (l+\frac{1}{2})\sum_{N=-J}^J \Lambda_N^J
     G_N^J(\mu )\int_0^\infty  d\omega  \omega^2
         (T_{l+\frac{1}{2}}(\omega)-T_{l+N+\frac{1}{2}}(\omega))^2.
\end{equation}
Here we used the formulae  (\ref{KG}) and (\ref{Gfunction}) for the
Clebsch Gordon coefficients. The evaluation of  $H(J)$   which is
quite involved, is given (together with all auxilary formulas )  in
the Appendices. The energy operator $H(J)$ has the following
expansion
\begin{equation}
   H(J)= \alpha_3 J^3+\alpha_2J^2 +\alpha_1 J+
   \frac{\alpha_{-1}}{J+\frac{1}{2}}+
   \frac{\alpha_{-2}}{(J+\frac{1}{2})^2}+\cdots.
\end{equation}
Terms with coefficients $\alpha_{-n}$,  $n=-1,-2,-3,\dots$  in the
above  expansion  give negligible  contributions to the Casimir
energy  compared to the first three ones
\begin{equation}\label{H}
   H(J)= - 3,03 J^3 - 3,37J^2-0,52J.
\end{equation}
The Casimir energy in the cavity obtained by the small but arbitrary
deformations of the  spherical region of radius $R$ is, therefore
given by
\begin{equation}\label{Casimir}
  E_{\tilde{S}}= \frac{\alpha}{R}( 1 + \frac{\beta}{4\pi R}\int d\Omega f+
   \frac{\beta^2}{4\pi R^2}\int d\Omega f(\hat{H} + 1)f ).
\end{equation}

\section{  Shape Dependence of the Casimir Energy, Discussion }
It is instructive to compare the Casimir energy (\ref{Casimir}) with
the energy in a spherical cavity with equal volume. The volume and
the area of the cavity after deformation are ( up to
$\frac{\beta^2}{R^2}$ order )
\begin{equation}\label{DefVolume}
   \tilde{ V}=\frac{4\pi}{3}R^3 (1-\frac{3\beta}{4\pi R}\int d\Omega
   f +\frac{3\beta^2}{4\pi R^2}\int d\Omega f^2 )
\end{equation}
and
\begin{equation}
   \tilde{S}=4\pi R^2 (1-\frac{\beta}{2\pi R}\int d\Omega
   f +\frac{\beta^2}{4\pi R^2}\int d\Omega f( 1-\frac{1}{2}\Delta )f )
\end{equation}
where
\begin{equation}
    \Delta =
    \frac{1}{\sin\theta}\frac{\partial}{\partial\theta}\sin\theta
    \frac{\partial}{\partial\theta}+\frac{1}{\sin^2\theta}\frac{\partial^2}{\partial\phi^2}
\end{equation}
is the Laplace operator on the sphere. The ratio of the  energy
(\ref{Casimir}) and the Casimir energy $E_0$ of the sphere with
volume (\ref{DefVolume}) is
\begin{equation}\label{ratioI}
    \frac{E_{\tilde{S}}}{E_0}|_{eq. \  vol.}=1+\frac{\beta^2}{4\pi
    R^2}(\int d\Omega f(2+\hat{H})f-\frac{1}{2\pi}(\int d\Omega f)^2
    ).
\end{equation}
This ratio to be examined by its dependence on the shape of the
cavity after deformation. One way may be to study its dependence on
the ratio of the surfaces of the deformed and spherical cavities
with equal volumes:
\begin{equation}
    \frac{\tilde{S}}{S_0}|_{eq. \  vol.}=1- \frac{\beta^2}{4\pi
    R^2}( \int d\Omega f(1+\frac{1}{2}\Delta )f-\frac{1}{4\pi}(\int d\Omega f)^2
    ).
\end{equation}
We then express the energy ratio (\ref{ratioI}) as
\begin{equation}
    \frac{E_{\tilde{S}}}{E_0}|_{eq. \  vol.}=1+\frac{\beta^2}{4\pi
    R^2}\int d\Omega f(\hat{H}-\Delta)f-2 (\frac{\tilde{S}}{S_0}|_{eq. \  vol.}-1)
\end{equation}
or
\begin{equation}\label{ratioII}
    \frac{E_{\tilde{S}}}{E_0}|_{eq. \  vol.}=1-
    2 \frac{\int d\Omega f(1+\frac{1}{2}\hat{H})f -\frac{1}{4\pi}(\int d\Omega f)^2}
    {\int d\Omega f(1+\frac{1}{2}\Delta)f-\frac{1}{4\pi}(\int d\Omega f)^2}
     (\frac{\tilde{S}}{S_0}|_{eq. \  vol.}-1).
\end{equation}
We know from (\ref{H}) that the operator $\hat{H}$ has negative
e-values. Thus both of the above relations show that the Casimir
energy linearly decreases by the increase of the surface. To have a
better feeling of this inverse proportionality  let us consider a
simple example. Suppose the deformation function is given by
\begin{equation}
    f(\theta,\phi)=P_l(\cos\theta)
\end{equation}
with $P_l$ being the Legendre polynomials. By using (\ref{Hoper})
and (\ref{CasIII}), we can write (\ref{ratioII}) for this specific
example as
\begin{equation}\label{ratioIII}
    \frac{E_{\tilde{S}}}{E_0}|_{eq. \  vol.}=1-
   \Lambda (l)(\frac{\tilde{S}}{S_0}|_{eq. \  vol.}-1)
\end{equation}
where the coefficient $\Lambda$ is given by
\begin{equation}
    \Lambda (l)=2\frac{2+H(l)}{2-l(l+1)}
\end{equation}
or by using (\ref{H}) can be written as
\begin{equation}
    \Lambda (l)=2\frac{2- 3,03 l^3 - 3,37l^2-0,52l}{2-l(l+1)}.
\end{equation}
\\
\\
Thus for large values of $l$ we can write (\ref{ratioIII}) as
\begin{equation}
    \frac{E_{\tilde{S}}}{E_0}|_{eq. \  vol.}=1-
   6l (\frac{\tilde{S}}{S_0}|_{eq. \  vol.}-1).
\end{equation}
We then can conclude that for less and less smooth deformations we
get smaller and smaller Casimir energies.

\begin{acknowledgments} The Authors thank the Turkish Academy of
Science (TUBA) for its supports,  to D.A. Demir and T.O. Turgut for
discussions and to P. Talazan for helping in "Mathematica".
\end{acknowledgments}

\appendix
\section{ The Calculation of the Vacuum Energy}

Using the uniform expansion formula (\ref{UNE}) for the spectral
function (\ref{SPF}) we represent the energy operator (\ref{Hoper})
( with $\mu=l+1/2$, $\omega=\mu x$ and $\varepsilon=\frac{N}{\mu}$ )
as
\begin{equation}
   H(J) =  \frac{1}{\pi \alpha} \sum_{n,m=0}^\infty
   \sum_{l=J}^\infty\mu^{2-2m-2n} \sum_{N=-J}^J \Lambda_N^J  D_l(N,J)
   \int_0^\infty  dx x^2   Y_{2n}(x,\varepsilon )Y_{2m}(x,\varepsilon )
\end{equation}
where
\begin{equation}
      Y_n (x,\varepsilon)=  \frac{d}{dx}(\frac{X_n(t(\varepsilon))}{(1+\varepsilon)^n}-X_n(t(0)))
\end{equation}
and
\begin{equation}
    t(\varepsilon) =\frac{1+\varepsilon}{\sqrt{(1+\varepsilon)^2+x^2}}.
\end{equation}
After the change of variables $2n=s+t$, $2m=s-t$ we have
\begin{equation}
   H(J) = \sum_{s=0}^\infty H^{s} (J)
\end{equation}
where
\begin{equation}
   H^s(J) =  \frac{1}{\pi \alpha}
   \sum_{l=J}^\infty\mu^{3-2s}\sum_{N=-J}^J \Lambda_N^J G_N^J(\mu)
   \int_0^\infty  dx x^2 F_s(x,\varepsilon)
\end{equation}
and
\begin{equation}
    F_s(x,\varepsilon)=\sum_{t=-s}^s  Y_{s+t}(x,\varepsilon )Y_{s-t}(x,\varepsilon
    ).
\end{equation}
By using the  Taylor expansion at $\varepsilon =0$
\begin{equation}
 F_s(x,\varepsilon)=\sum_{k=0}^\infty \frac{\varepsilon^n}{n!}F^{(n)}_sx)
\end{equation}
 and  the asymptotic expansion (\ref{asymptotic}) we arrive at
\begin{equation}\label{PiII}
   H^s(J) = \sum_{\tau=0}^\infty b_\tau \zeta
   (2\tau+2s-2, J)
\end{equation}
where
\begin{equation}
\zeta(z,J)=\sum_{k=0}^\infty\frac{1}{(J+\frac{1}{2}+k)^z}
\end{equation}
is the Riemann zeta function
\begin{equation}
    b_\tau =\frac{1}{\pi \alpha} \sum_{p=-\tau}^\tau \frac{\langle Z_{\tau+p}
    N^{\tau-p}\rangle}{(\tau-p)!}\int_0^\infty dx x^2
    F^{(\tau-p)}_s(x)
\end{equation}
and
\begin{equation}
    \langle f(N) \rangle=\sum_{N=-J}^J \Lambda_N^J f(N).
\end{equation}
For large $J$ we have
\begin{equation}
   \zeta    (2\tau+2s-2, J)\simeq J^{3-2\tau-2s}
\end{equation}
and
\begin{equation}
    \langle Z_{\tau+p}
    N^{\tau-p} \rangle\simeq J^{2\tau}.
\end{equation}
This asymptotic expressions imply
\begin{equation}
  H^s(J)\simeq J^{3-2s}, \ \ \  J\gg 1.
\end{equation}
Thus $H^{0}(J)$ and $H^1(J)$ give the main contributions to the
energy operator $H(J)$. Using the summation formulae
(\ref{summation}) we  get
\begin{equation}
   H^{0} (J) = \frac{1}{128\alpha}(
  J(J+1) \zeta( 0,J) - \frac{J(J+2)(J^2-1)}{8}\zeta( 2,J))
\end{equation}
and
\begin{equation}
   H^{1} (J) =-\frac{9}{ 2048\alpha}(
   \frac{J(J+1)}{2} \zeta( 2,J) + \frac{99}{16\cdot 24}(J+\frac{1}{2})^4 \zeta(
   4,J))
\end{equation}
 from which by the virtue of
\begin{eqnarray}
  \zeta (2,J) &\approx& \frac{1}{J+1/2}+\frac{1}{2(J+1/2)^2}+\frac{1}{6(J+1/2)^3}+o(J) \nonumber \\
\zeta(4,J) &\approx& \frac{1}{3(J+1/2)^3}+\frac{1}{2(J+1/2)^4} +
o(J)
\end{eqnarray}
we read
\begin{equation}
\alpha_{3}=- \frac{9}{1024\alpha}, \ \ \ \alpha_{2}=-
\frac{10}{1024\alpha}, \ \ \ \alpha_{1}=- \frac{1193}{1024\cdot
768\alpha}
   \end{equation}

\section{ The Klebsch Gordon coefficients}
The spherical harmonics are
\begin{equation}\label{sphharm}
    Y^l_m(\theta,\phi)=\{
    \begin{array}{c}
    2 P^m_l(\cos\theta )\cos m\phi, \ \ \  m=1,2,\dots l  \\
    P_l(\cos\theta ), \ \ \  m=0 \\
    2 P^{\mid m\mid} _l(\cos\theta ) \sin \mid m\mid\phi, \ \ \   m=-1,-2,\dots -l
    \end{array}
\end{equation}
where $P_l(x)$ is the Legendre function and
\begin{equation}
    P^m_l(x )=\sqrt{\frac{l-m)!}{(l+m)!}}(1-x^2)^{\frac{m}{2}}\frac{d^m}{dx^m}P_l(x)
\end{equation}
is the associated Legendre function. We use the notation
$Y^l_m(\theta,\phi )=Y^l_m(\vec{n})$ where
\begin{equation}\label{unitvector}
  \vec{n}=(\cos\theta, \sin\theta\sin\phi, \sin\theta\cos\phi )
\end{equation}
is the unit vector on the sphere.

The  addition formula for the spherical harmonics is
\begin{equation}\label{add}
    P_l((\vec{n}, \vec{n^\prime}) )=\sum_{m=-l}^l   Y^l_m(\hat{n}) Y^l_m(\hat{n}^\prime)
\end{equation}
where the argument of the Legendre function is the scalar product of
the unit vectors $\vec{n}$ and $\vec{n}^\prime$.

The Klebsch Gordon coefficients $K^J_{ll+N}$  defined by the
expansion
\begin{equation}\label{expansion}
    P_l(x)P_{l^\prime}(x)= \sum_{J=\mid l-l^\prime\mid}^{l+l^\prime}K^J_{ll^\prime} P_J (x)
\end{equation}
are  \cite{vil}
\begin{equation}\label{KG}
    K^J_{ll+N} = (J+\frac{1}{2})
    \Lambda_N^J F_l(N,J)
\end{equation}
where ( $\mu=l+1/2$ )
\begin{equation}
    F_l(N,J)=\frac{1}{\mu+\frac{N+J}{2}}\frac{ \Gamma (\mu+ \frac{N-J}{2})\Gamma (\mu +\frac{N+J}{2}+\frac{1}{2})}
   {    \Gamma (\mu+ \frac{N-J}{2}+\frac{1}{2})    \Gamma (\mu+\frac{N +J}{2})}
\end{equation}
and
\begin{equation}
   \Lambda_N^J=\frac{1}{\pi}\frac{\Gamma (\frac{J-N+1}{2})\Gamma (\frac{J+N+1}{2})}
   {\Gamma (\frac{J-N+2}{2})\Gamma (\frac{J+N+2}{2})}
\end{equation}
These coefficients  are  nonzero if  $\mid N \mid  \leq J\leq 2l+N$
and   $N+J$ is even number.

The formula
\begin{equation}
   \ln \Gamma (z)=z\ln z -z -\frac{1}{2}\ln z+\ln\sqrt{2\pi}+
   \sum_{k=1}^{n-1}\frac{B_{2k}}{2k(2k-1)z^{2k-1}}
\end{equation}
implies
\begin{equation}
   \ln\frac{\Gamma (z+\frac{1}{2}}{\Gamma (z)}=\frac{1}{2}\ln z -\frac{1}{8z}+\frac{1}{192 z^3}
\end{equation}
up to the third order in $\frac{1}{z}$. The last formula allows as
to get the asymptotic expansion of the functions $F_l(N,J)$. Up to
the third order in $\frac{1}{\mu}$ for large values of $\mu$ we
have:
\begin{equation}
F_l(N,J)=\frac{1}{\sqrt{(\mu+\frac{N-J}{2})
(\mu+\frac{N+J}{2})}}(1+\frac{J}{8\mu^2}-\frac{NJ}{8\mu^3})
\end{equation}
For  the function
\begin{equation}\label{Gfunction}
   G_N^J(\mu )= (\mu+N)F_l(N,J)
\end{equation}
we get the following asymptotic expansion
\begin{equation}\label{asymptotic}
    G_N^J(\mu ) = 1+ \frac{Z_1}{\mu}+ \frac{Z_2}{\mu^2}+  \frac{Z_3}{\mu^3}+
    \cdots
\end{equation}
where
\begin{eqnarray}
    Z_1 &=& \frac{N}{2}, \nonumber \\
    Z_2 &=& \frac{J(J+1)-2N^2}{8}, \nonumber  \\
    Z_3 &=& N \frac{2N^2-J(J+1)}{16}.
\end{eqnarray}
\\
\\
Let  $J$ is even. Putting $J=2j$ and $N=2n$ we have
\begin{equation}
   \sum_{N=-J}^{J}\Lambda_N^Je^{iN\theta}=
   \sum_{n=-j}^{j}\frac{1}{\pi}\frac{\Gamma (j-n+\frac{1}{2})\Gamma (j+n+\frac{1}{2})}
   {(j-n)! (j+n)!}e^{i2n\theta}
\end{equation}
In the new variable $m=n+j$ we have
\begin{equation}
   \sum_{N=-J}^{J}\Lambda_N^Je^{iN\theta}=
   \sum_{m=0}^{2j}\frac{1}{\pi}\frac{\Gamma (m+\frac{1}{2}\Gamma (2j-m+\frac{1}{2})}
   {m! (2j-m)! }e^{i(2m-2j)\theta}
\end{equation}
which is exactly the series representation for zonal spherical
functions $Y^{2j}_0(\cos\theta )$
\begin{equation}
   \sum_{N=-J}^{J}\Lambda_N^Je^{iN\theta}=Y^J_0(\cos\theta ).
\end{equation}
The same formula is true for odd $J$. For example we have
\begin{eqnarray}\label{summation}
   \sum_{N=-J}^{J}\Lambda_N^J &=& 1, \nonumber \\
   \sum_{N=-J}^{J}\Lambda_N^JN^2 &=&  \frac{J(J+1)}{2}, \nonumber \\
   \sum_{N=-J}^{J}\Lambda_N^JN^4 &=& \frac{3J^3(J+2)+J(J-2)}{8}.
\end{eqnarray}

\end{document}